\documentclass[letterpaper, 10 pt, conference]{ieeeconf}  % Comment this line out
\IEEEoverridecommandlockouts                              % This command is only
\makeatletter
\let\NAT@parse\undefined
\makeatother

\usepackage[dvipsnames]{xcolor}

\newcommand*\linkcolours{ForestGreen}

\usepackage{times}
\usepackage{graphicx}
\usepackage{amssymb}
\usepackage{gensymb}
\usepackage{amsmath}
\usepackage{breakurl}

\usepackage{url,hyperref}
\hypersetup{
colorlinks,
linkcolor=\linkcolours,
citecolor=\linkcolours,
filecolor=\linkcolours,
urlcolor=\linkcolours}

\usepackage{algorithm}
\usepackage{algorithmic}

\usepackage[labelfont={bf},font=small]{caption}
\usepackage[none]{hyphenat}

\usepackage{mathtools, cuted}

\usepackage[noadjust, nobreak]{cite}
\usepackage{tabularx}
\usepackage{amsmath}

\usepackage{float}

\usepackage{pifont}% http://ctan.org/pkg/pifont

\newcolumntype{Y}{>{\centering\arraybackslash}X}

\usepackage[]{placeins}

\usepackage{placeins}

\usepackage{tikz}

\usepackage[framemethod=tikz]{mdframed}

\usepackage{afterpage}

\usepackage{stfloats}

\usepackage{atbegshi}
\newcommand{\handlethispage}{}
\newcommand{\discardpagesfromhere}{\let\handlethispage\AtBeginShipoutDiscard}
\newcommand{\keeppagesfromhere}{\let\handlethispage\relax}
\AtBeginShipout{\handlethispage}

\usepackage{comment}
\usepackage[normalem]{ulem}

\usepackage{setspace}
\usepackage{authblk}
\title{\LARGE \bf
Efficient partitioning of surface Green’s function: toward \textbf{\textit{ab initio}} contact resistance study.
}

\author[1,2]{G. Gandus}
\author[2]{Y. Lee}
\author[1]{D. Passerone}
\author[2]{M. Luisier}
\affil[1]{nanotech@surfaces (EMPA)}
\affil[2]{Integrated Systems Laboratory (ETH Zurich)}

\newcommand{\bb}[1]{\boldsymbol{#1}}

\begin{document}

\maketitle
\thispagestyle{empty}
\pagestyle{empty}

\begin{abstract}

%In this work, we propose an efficient computational scheme based on the non-equilibrium Green's function (NEGF) formalism to evaluate the open-boundary conditions for first-principles quantum transport simulations of quasi 1D channels connected to bulk electrodes. In conventional methods, the self-energy matrices of the leads are computed for a supercell which is restricted to have the same lateral dimensions of the adjoining atoms in a scattering region. Here, we obtain the properties of the bulk electrodes from the calculation of the Green's function for smaller cells that compose the large supercell. Partitioning the contacts into smaller building blocks leads to significant improvements in the computational efficiency of electron-transport calculations without sacrificing the accuracy of the results. To exemplify the merits of the proposed method we investigate the contact resistance carrier density dependency in silicon nanowire devices with metallic contacts. 

In this work, we propose an efficient computational scheme for first-principle quantum transport simulations to evaluate the open-boundary conditions. Its partitioning differentiates from conventional methods in that the contact self-energy matrices are constructed on smaller building blocks, principal layers (PL), while conventionally it was restricted to have the same lateral dimensions of the adjoining atoms in a channel region. Here, we obtain the properties of bulk electrodes through non-equilibrium Green's function (NEGF) approach with significant improvements in the computational efficiency without sacrificing the accuracy of results. To exemplify the merits of the proposed method we investigate the carrier density dependency of contact resistances in silicon nanowire devices connected to bulk metallic contacts.

\end{abstract}

%%%%%%%%%%%%%%%%%%%%%%%%%%%%%%%%%%%%%%%%%%%%%%%%%%%%%%%%%%%%%%%%%%%%%%%%%%%%%%%%
\section{INTRODUCTION}
\begin{figure}[b]
    \begin{center}
        \includegraphics[width=0.45\textwidth]{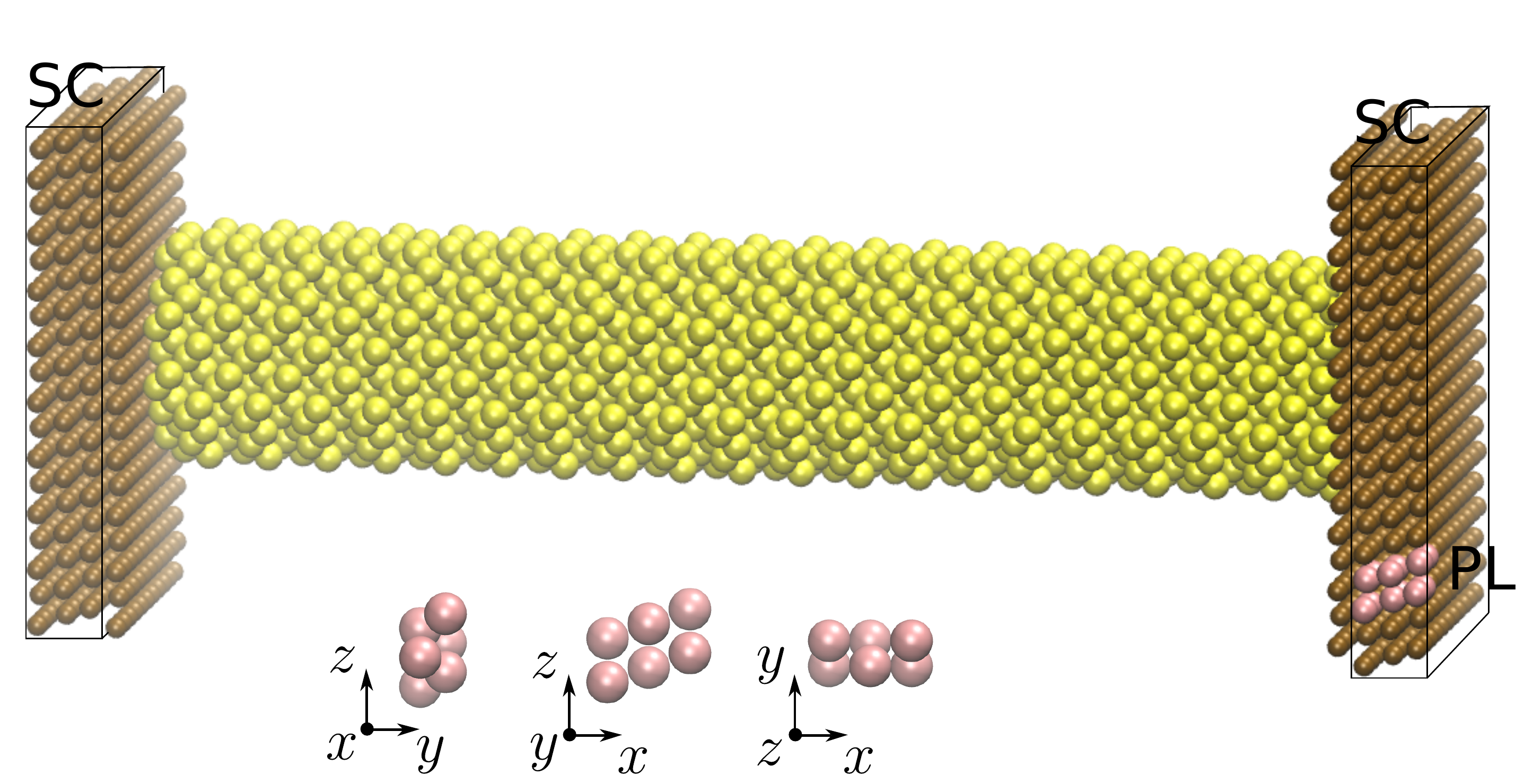}
    \end{center}
    \caption{Schematic view of a circular Silicon nanowire (SiNW) structure with bulk-like metallic contacts surrounding it. Yellow spheres represent Si and H atoms (9 and 5 orbitals per atom), the brown ones Au (9 orbitals per atom). The black rectangles mark the SC contacts, while the pink spheres highlight one PL composing the SC contacts. The latter is constructed by repeating a single PL in the transverse directions perpendicular to the nanowire axis.}
    \label{fig:Device}
\end{figure}
In recent years, quasi one-dimensional (1D) materials such as nanowires and nanotubes have been at the forefront of nanotechnology research. Their unique electronic properties make them promising candidates as next-generation logic switches. However their performance, is often limited by their high contact resistance~\cite{bourdet2016contact}. 
%Simulation of models as close as possible to reality has been indispensable for designing the most promising contact prototypes for nanowire devices. In this regard, first-principle quantum transport simulations have become the standard to accurately evaluate the transport properties between the metal contact and devices of interest, within the atomic precision. 
First-principles quantum transport simulations are ideally suited to design contacts with low resistances based on atomistic models as close as possible to reality.
Such \textit{ab initio} investigations require heavy computational burden as compared to, e.g., tight-binding or k$\cdot$p models. One of the major bottlenecks arises from calculating the boundary conditions. 
%More precisely, the self-energy matrices must be computed for a supercell (SC)-sized contact that is constrained to have the same lateral extensions of the adjoining atoms as in the neighbor scattering region. 
Usually, the electrodes employed in 1D devices are bulk like, i.e. they consist of a repeating unit cell. The standard method in transport calculations is to build a supercell (SC)-sized contact which is constraint to have at least the same lateral dimensions as the channel (see Fig.~\ref{fig:Device}). Because the computational cost to evaluate the boundary self-energies is in general proportional to the cube of the system size, large-scale electron-transport characterization are often impractical.
%Usually, however, the contacts are finite blocks of bulk materials that consists of a simple repetition of unit cells. Band structure theory predicts that the electronic information of systems that can be break down into a stack of unit cells, is in principle fully contained in the properties of the elementary blocks. Therefore, one needs not to consider explicitly the atomic structure of the entire contacts. On the basis of this observation, we propose a method to obtain the self-energy matrix for a SC electrode from the Green's function of a smaller unit: the principal layer.}  

In this work, we propose an efficient algorithm to address this issue, partitioning the contact self-energy into smaller building blocks called principal layers (PL). This method allows to fully retain the electronic information of the contacts, while significantly improving the computational efficiency, on the basis of Bloch's theorem. As an application, we demonstrate the strength of the proposed approach for large-scale quantum transport simulations of silicon nanowires with metallic electrodes and report the resulting contact resistances as a function of the nanowire diameters and carrier densities. 

%%%%%%%%%%%%%%%%%%%%%%%%%%%%%%%%%%%%%%%%%%%%%%%%%%%%%%%%%%%%%%%%%%%%%%%%%%%%%%%%
\section{ALGORITHM}

\begin{figure}[h]
    \begin{center}
        \includegraphics[width=0.45\textwidth]{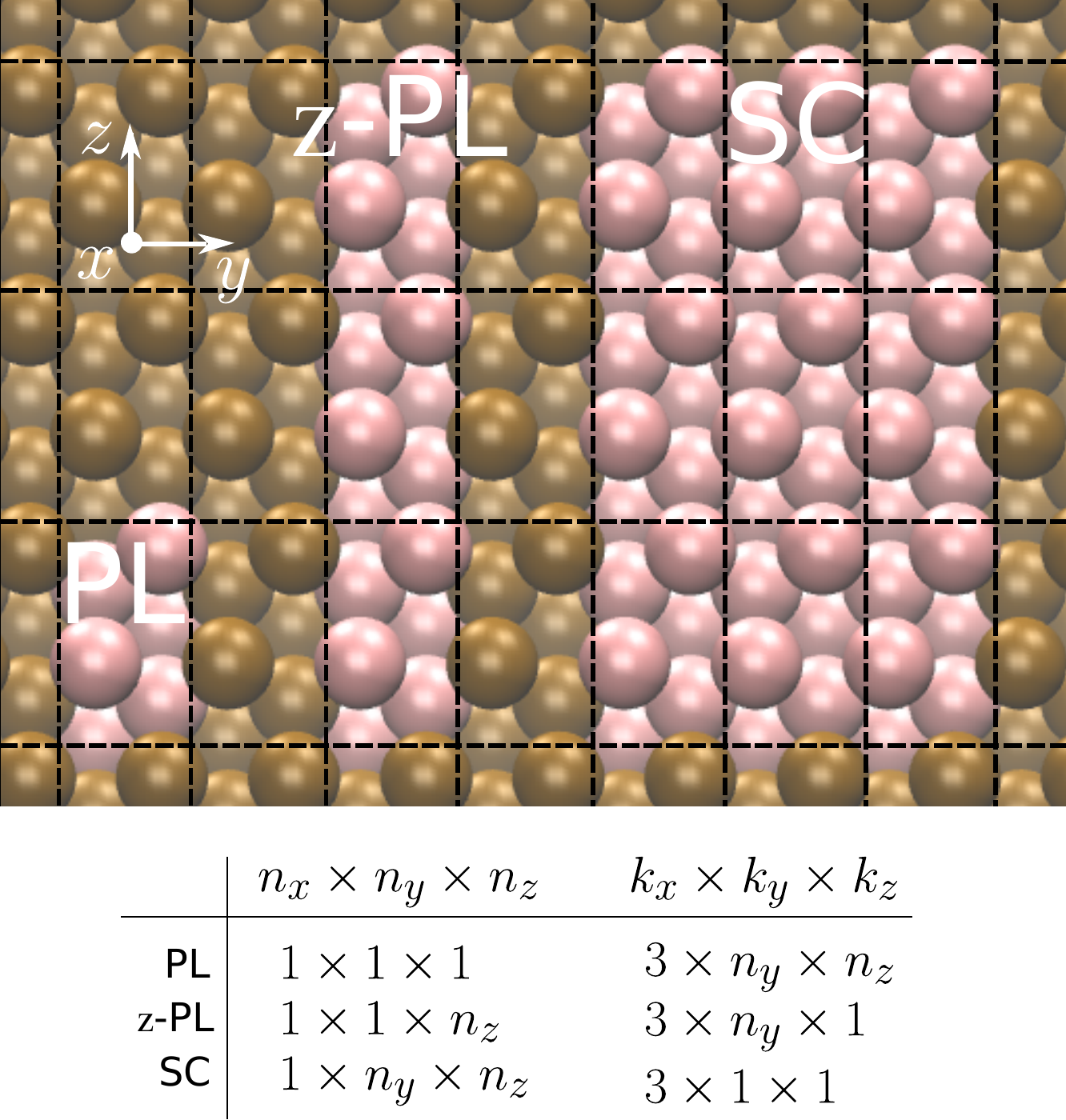}
    \end{center}
    \caption{Schematic representation of a PL, a $z$-PL and a SC on Au(111) surface. The table summarizes the number of PL repetitions in real space (left column) and the corresponding k-space mesh (right column) used in the electronic structure calculation to ensure a correct inclusion of the minimum image convention.}
    \label{fig:Algorithm}
\end{figure}

Our self-energy partitioning algorithm is summarized in Fig.~\ref{fig:Algorithm}. We consider a SC contact as a stack of principal layers (PLs) that are repeated $n_y \times n_z $ times along the $y$- and $z$-directions transverse to the transport $x$-direction. 
We assume that: i) the PL has only nearest neighbor couplings ($n_x=1$) in the $x$-direction, with $n_y$ and $n_z$ being odd numbers; ii) the SC is periodic in the $y$- and $z$-directions.
First, the electronic structure of the periodic PL is computed with a local basis set of orbitals. The Brillouin zone is sampled with a $3 \times n_y \times n_z$ k-space grid. This ensures
%that the periodic repetitions are embedded in a SC with the same dimensions of the realspace structure and the couplings at the SC boundaries are treated on an equal footing.
that projecting the SC into the PL leads to the corresponding Bloch-space Hamiltonian.
We then carry out a partial Bloch sum over the 3 wavevectors in the $x$-direction,
\begin{equation}\label{eq:H0}
    \bb{H}_0(\bb{k}_{t}) = \sum_{k_{x}} \bb{H}(\bb{k}) e^{i(k_x,\bb{k}_{t}) \cdot (0,0,0)}, \\
\end{equation}
\begin{equation}\label{eq:H1}
    \bb{H}_1(\bb{k}_{t}) = \sum_{k_{x}} \bb{H}(\bb{k}) e^{i(k_x,\bb{k}_{t}) \cdot (1,0,0)},
\end{equation}
where $\bb{k}_t$ includes the transverse components ($k_y$ and $k_z$) of $\bb{k}$. The $\bb{H}_i(\bb{k})$ are the $k$-dependent Hamiltonian matrices. The overlap matrices $\bb{S}_0(\bb{k}_{t})$ and $\bb{S}_1(\bb{k}_{t})$ are obtained in the same way. Eqs.~(\ref{eq:H0}) and (\ref{eq:H1}) transform a set of fully-delocalized Bloch wavefunctions into a set of hybrid wave-functions which remain extended (Bloch-like) along the transverse directions, but are localized in the transport direction. We note that the aforementioned transformation enables the usage of an iterative scheme~\cite{sancho1984quick} to compute the surface Green's function $\bb{G}(E,\bb{k}_t)$. Hereafter, we drop the energy dependence $E$ for ease of readability. By performing a Bloch summation over the transverse wavevectors, the distance-dependent Green's functions $\bb{G}_{n,m}$ can be finally obtained as
\begin{equation}\label{eq:Gnm}
    \bb{G}_{n,m} = \sum_{\bb{k}_{t}} \bb{G}(\bb{k}_{t}) e^{i(\bb{k}_{t}) \cdot (n,m)},
\end{equation}
where $n,m$ refer to the position of the PL within the SC. The surface Green's function within one PL is simply given by $\sum_{\bb{k}_{t}} \bb{G}(\bb{k}_{t})$. As last step, we construct the SC Green's function $\bb{G}$ as the circulant matrix 
%shown in Fig.~\ref{fig:Algorithm}.
\begin{equation} \label{eq:Gs}
    \bb{G} = \begin{bmatrix}
             \bb{G}^z_{0} & \bb{G}^z_{1} & \cdots & \bb{G}^z_{-1} \\
             \bb{G}^z_{-1} & \bb{G}^z_{0} & \cdots & \bb{G}^z_{1} \\
             \vdots &  & \ddots & \vdots \\
             \bb{G}^z_{1} & \bb{G}^z_{-1} & \cdots & \bb{G}^z_{0} \\
             \end{bmatrix},
\end{equation}
with matrix elements that are themselves circulant
\begin{equation} \label{eq:Gz}
    \bb{G}^z_i = \begin{bmatrix}
             \bb{G}_{i,0}  & \cdots & \bb{G}_{i,-1} \\
             \vdots &  \ddots & \vdots \\
             \bb{G}_{i,1} & \cdots & \bb{G}_{i,0} \\
             \end{bmatrix}.
\end{equation}
The indices of the first row in $\bb{G}$ and $\bb{G}^z_\text{i}$ are: $[0, 1, ..., (n_y-1)/2, -(n_y-1)/2, ..., -1]$ and $[0, 1, ..., (n_z-1)/2, -(n_z-1)/2, ..., -1]$, respectively. 
The boundary self-energies can then be derived from the matrix products~\cite{thygesen2006electron}
\begin{equation} \label{eq:Sigmaz}
    \bb{\Sigma}(z) = (z\bb{S}_1^{\dag}-\bb{H}_1^{\dag})\bb{G}(z)(z\bb{S}_1-\bb{H}_1),
\end{equation}
where $z=i\eta+E$ with $\eta$ a small positive number while $\bb{H}_1$ and $\bb{S}_1$ are the Hamiltonian and overlap matrices, respectively, connecting two consecutive SC in the $x$-direction. They are evaluated in the same way as $\bb{G}$ from the elements of Eq.~(\ref{eq:H1}).

%Here, we highlight that this partitioning, thanks to the odd condition on $n_y$ and $n_z$, leads to $(n-1)/2+1$ degrees of freedom rather than $n^2$, while ensuring a correct inclusion of the minimum-image convention.  

Here we highlight the computational efficiency of the proposed algorithm. Indeed, the iterative scheme to compute the surface Green's function requires to calculate the inverse of a matrix with the same dimensions as $\bb{G}$. The computational cost is then $O((nN)^3)$, where $n=n_y \times n_z$ and $N$ is the number of basis functions in one PL. Thanks to the proposed partitioning, the computational complexity of the recursive algorithm is reduced to $O(nN^3)$ as one has to invert $n$ distance-dependent Green's functions $\bb{G}_{n,m}$, each of dimension $N$. This scheme is therefore particularly advantageous to simulate 1D devices with large channels.
We also note that it is the odd condition on $n_y$ and $n_z$ that ensures a correct inclusion of the minimum-image convention. Careful attention must be payed when constructing the SC contacts which must be able to be partitioned into an odd number of PLs. As an alternative, one can consider a stack of PLs as the smallest building block for which the distant-dependent Green's functions are evaluated. This is the case when either $n_y$ or $n_z$ is an even number.

%%%%%%%%%%%%%%%%%%%%%%%%%%%%%%%%%%%%%%%%%%%%%%%%%%%%%%%%%%%%%%%%%%%%%%%%%%%%%%%%
\section{RESULTS}
\begin{figure}[h]
    \begin{center}
        \includegraphics[width=0.45\textwidth]{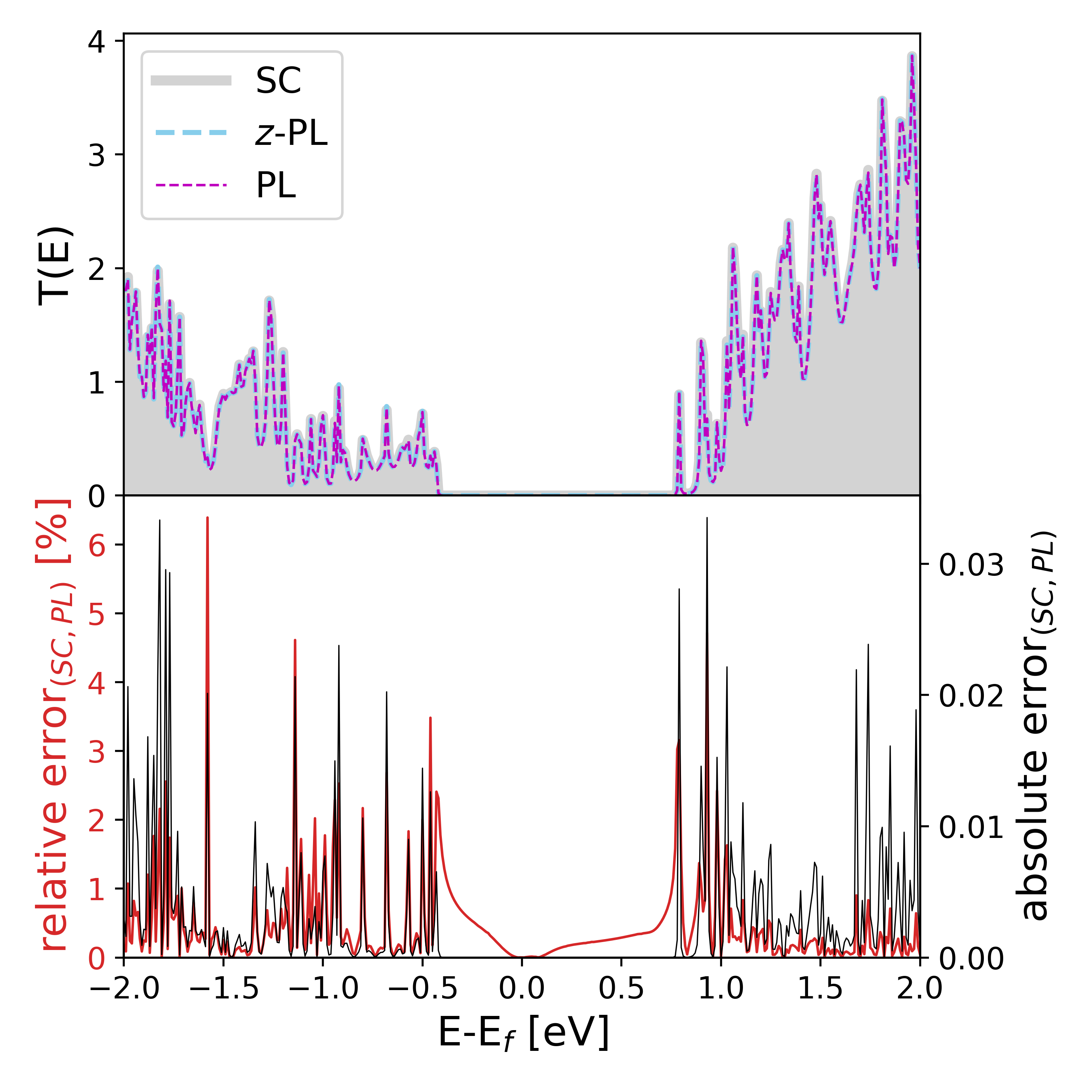}
    \end{center}
    \caption{(Top panel) Conductance spectra of a SiNW obtained with the entire SC, a $z$-PL, and a single PL. The nanowire has a length of $10$ nm and a diameter of $2$ nm. (Bottom panel) numerical difference between the SC and PL transmissions.}
    \label{fig:Conductance}
\end{figure}

As benchmark example, we consider a Silicon nanowire (SiNW) attached to bulk Au(111) surfaces~\cite{mohney2005measuring}.
%{\color{red}{\texbf{, as shown in Fig}}}. 
The SC are modeled by a three-layer-thick Au(111) and are composed of $7 \times 5$ PLs, each having $6$ atoms with $9$ orbitals per atom. The scattering region includes the nanowire extended to each terminal with four Au(111) slabs (see Fig.~\ref{fig:Device}). The $H$ and $S$ matrices are computed from density functional theory (DFT) with GPAW \cite{enkovaara2010electronic} using single zeta polarized (SZP) basis functions and the Perdew, Burke, Ernzerhof (PBE) exchange-correlation functional.
To validate the accuracy of our approach, we compare the conductance spectra obtained with the entire SC, a stack of PLs with $n_y \times n_z = 1 \times 5$ ($z$-PL), and a single PL. Fig.~\ref{fig:Conductance} shows that
the proposed algorithm successfully reproduces the SC conductance spectra within $3\times10^{-2}$ absolute errors.

\begin{figure}[h]
    \begin{center}
        \includegraphics[width=0.45\textwidth]{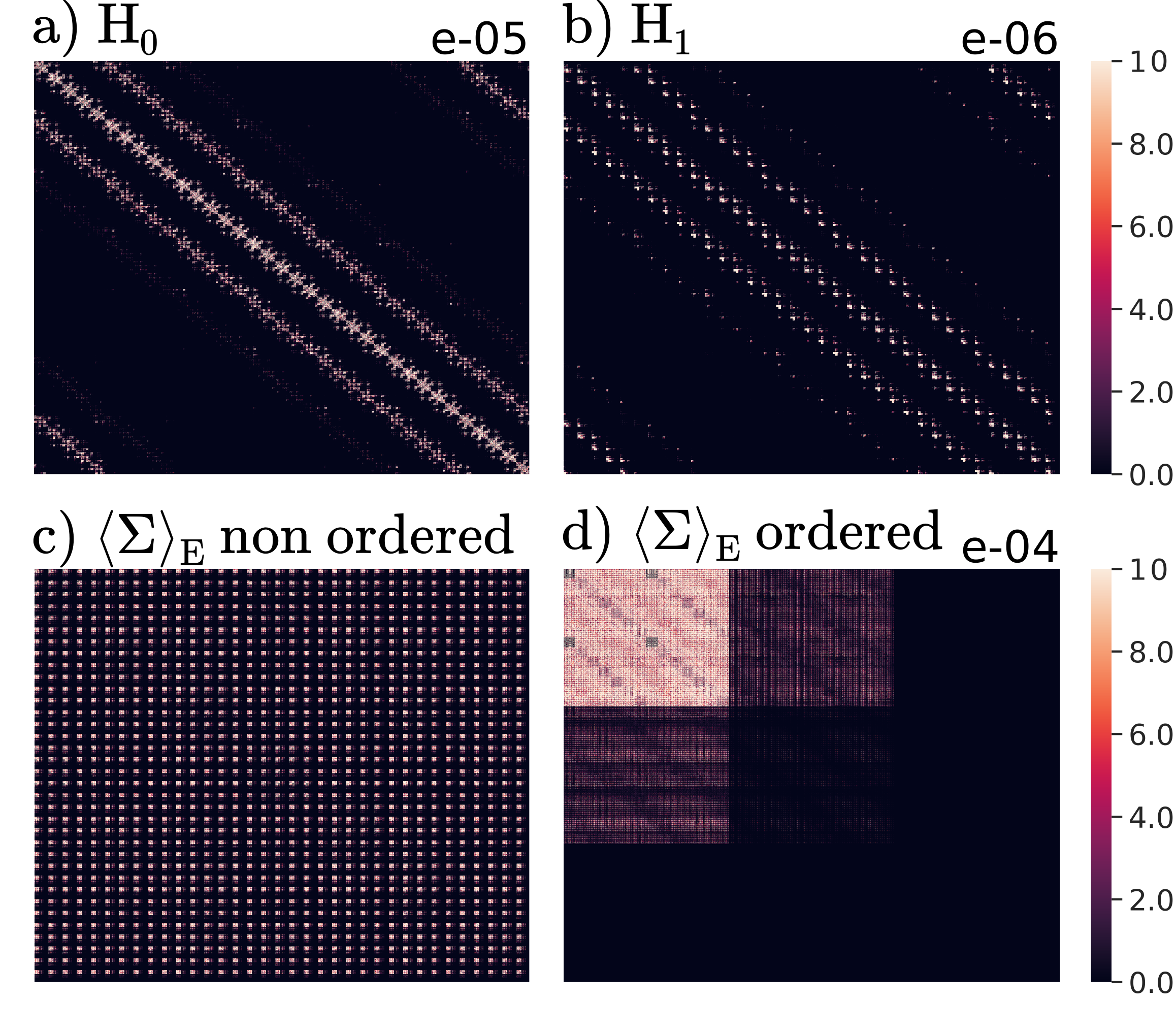}
    \end{center}
    \caption{Absolute error in (a) $\bb{H}_0$ and (b) $\bb{H}_1$ between the Hamiltonian matrices obtained by an electronic structure calculation of a single PL with respect to the reference SC. (c) Absolute error for each entry in $\bb{\Sigma}(z)$ averaged over an energy range from $-4$eV to $4$eV with steps of $0.5$eV. (d) Same as (c), but with matrix elements arranged in such a way that the SC atoms are ordered along the transport direction.}
    \label{fig:H01_Sig}
\end{figure}

To investigate the influence of the discrepancy in electronic structure results on the conductance spectra, we compare the $\bb{H}_0$ and $\bb{H}_1$ matrices obtained by the SC and PL approaches. For the former, the matrices are directly obtained by Eqs.~(\ref{eq:H0}) and (\ref{eq:H1}) since the wavefunctions are already real in the transverse directions, whereas for the latter, they are constructed following the same steps outlined for $\bb{G}$ Eqs.~(\ref{eq:H0}) to (\ref{eq:Gz}). Figs.~\ref{fig:H01_Sig}(a-b) report the absolute errors for each entry of the matrices. 
As expected, the large errors appear in the entries connecting neighboring PLs and are comparable with the average absolute error in conductance (see black line in the bottom panel of Fig.~\ref{fig:Conductance}). 
We conclude that it is already at the DFT level that the majority of the errors are accumulated.
When constructing $\bb{H}_0$ with Eqs.~(\ref{eq:Gs}) and (\ref{eq:Gz}) we assume that the electronic potential in the PLs is incommensurate with the SC structure, i.e., the diagonal entries $\bb{H}^z_0$ and $\bb{H}_{i,0}$ are equivalent. However, the Hamiltonian of the SC does not reflect this symmetry.

To see how the discrepancies introduced by the DFT calculation propagate in the recursive algorithm for computing the surface Green's function, we evaluate the absolute error between the boundary self-energies obtained with the Hamiltonian and overlap matrices corresponding to the PL and SC structures. To achieve this without loss of generality for energy points, we compute the average error for each entry in $\bb{\Sigma}(z)$ using $16$ energy points in the range from $-4$eV to $4$eV with a regular grid spacing of $0.5$eV.
The result is illustrated in Fig.~\ref{fig:H01_Sig}(c) where it can be observed that the absolute error overall increases by one order of magnitude when compared to Figs.~\ref{fig:H01_Sig}(a-b). This is attributed to the average number of iterations required to converge $\bb{G}$, which is found to be $~12$. The error adds up linearly at each iteration step. 
To gain further insights into the distribution of the error, we reorder the matrix elements from Fig.~\ref{fig:H01_Sig}(c) so that the atoms in the SC structure are aligned along the transport direction~\footnote{The reordering is applied consistently in the conductance calculations where the adjoining atoms in the scattering region are ordered for increasing $x$-coordinate so as to maintain a block tridiagonal structure of the scattering Hamiltonian~\cite{dmitrytheory}.}.
The resulting matrix is plotted in Fig.~\ref{fig:H01_Sig}(d). It can be seen that the majority of the errors appear in the entries in the upper-left corner which couples the electrodes to the adjoining atoms in the scattering region closer to the interface.  

\begin{figure}[h]
    \begin{center}
        \includegraphics[width=0.45\textwidth]{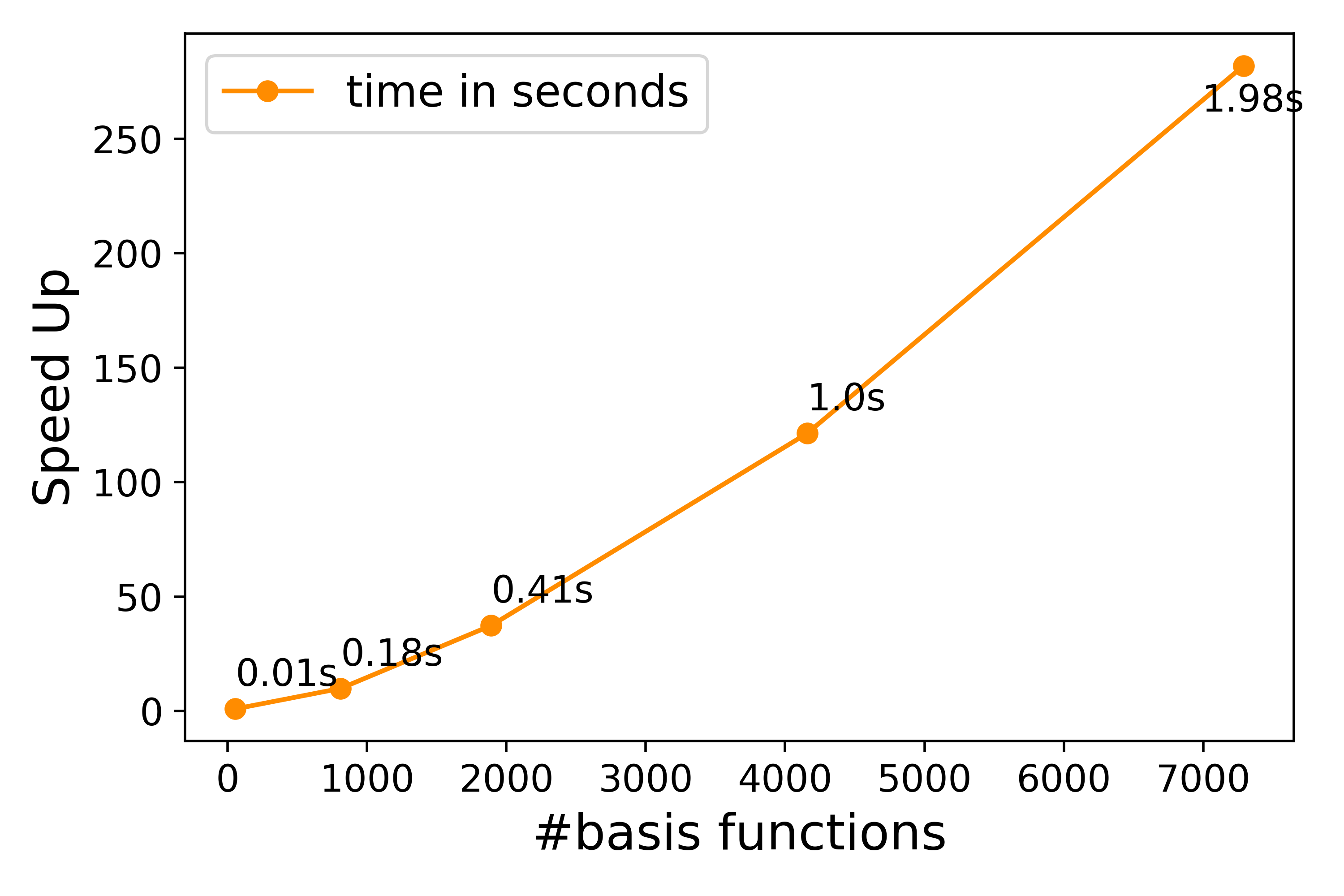}
    \end{center}
    \caption{Scalability tests to generate the self-energy matrix of a Au-SiNW-Au structure as a function of the number of basis functions. The data points refer to SCs with $n_y\times n_z =$ $1\times1$, $5\times3$, $7\times5$, $11\times7$ and $15\times9$. The speed up factor is obtained as [SC time / PL time]. In addition, the CPU time required by the proposed algorithm is reported on top of the data points.}
    \label{fig:SpeedUp}
\end{figure}

The computational efficiency of our approach is summarized in Fig.~\ref{fig:SpeedUp} where we report the SC/PL Speed Up in generating the self-energies terms for several SC sizes. The result shows that the CPU time could be accelerated by a factor 280 for the largest SC, reducing the time from $554.43$ to $1.98$ sec.

\begin{figure*}[ht]
    \begin{center}
        \includegraphics[width=1.\textwidth]{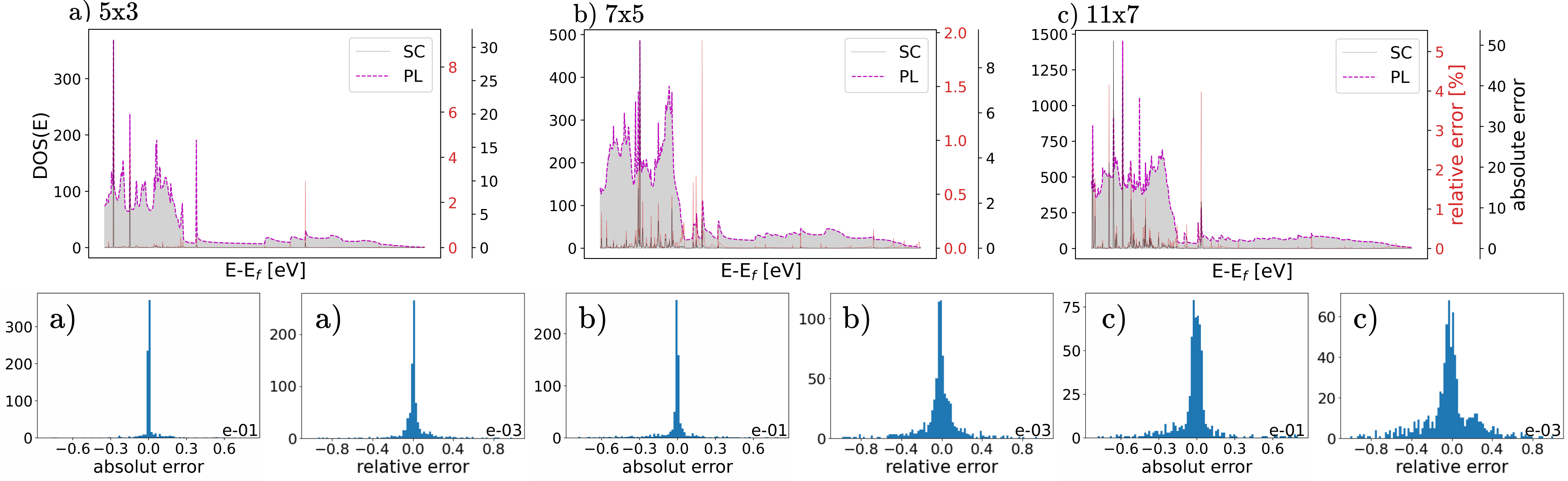}
    \end{center}
    \caption{(Top panel) Energy-resolved DOS of the device contacts in Fig.~\ref{fig:Device} (left axis) and absolute and relative errors between the proposed method and the reference SC (right axis). (Bottom panel) Histogram of the error for the considered contact models.} 
    \label{fig:DOS}
\end{figure*}

To verify that increasing the SC size does not deteriorate the accuracy, we evaluate the density of states (DOS) for a selection of SC sizes. The DOS is given by the formula:  
\begin{equation} \label{eq:DOS}
    DOS(E) = Tr \left[ \bb{G}(E) \bb{S}_0 \right].
\end{equation}
From Eq.~(\ref{eq:DOS}) we can see that this quantity includes both diagonal and off diagonal errors into one number. The energy-resolved DOS is shown in the top panel of Fig.~\ref{fig:DOS}(a-c) for varying $n_y \times n_z$. In all considered cases, the DOS obtained by a single PL agrees almost perfectly with the SC reference, thus confirming the strength and validity of our scheme. 
The absolute and relative errors (with the error-axis on the right side of top panel in Fig.~\ref{fig:DOS}(a-c)) are overlaid on the DOS curves. Both errors are more pronounced at the position of the peaks in the DOS curves. From the distribution of the error summarized in the histograms in the bottom panel of Figs.~\ref{fig:DOS}(a-c) we observe that the variance increases with the SC size. However, the maximum error does not show a SC-size dependence as can be seen in Table~\ref{table:1}.

\begin{table}[h]
\centering
\begin{tabular}{ |c||c|c|  }
 \hline
 \multicolumn{3}{|c|}{Maximum error} \\
 \hline
 $n_y \times n_z$&absolute error&relative error\\
 \hline
 5x3&31.13&0.092\\
 7x5&9.22&0.019\\
 11x7&42.21&0.04\\
 \hline
\end{tabular}
\caption{Table summarizing the maximum absolute and relative errors found for the DOS curves in Fig.~\ref{fig:DOS}. All maximum errors are found correspond to peaks in the DOS curves.}
\label{table:1}
\end{table}

\begin{figure}[h]
    \begin{center}
        \includegraphics[width=0.45\textwidth]{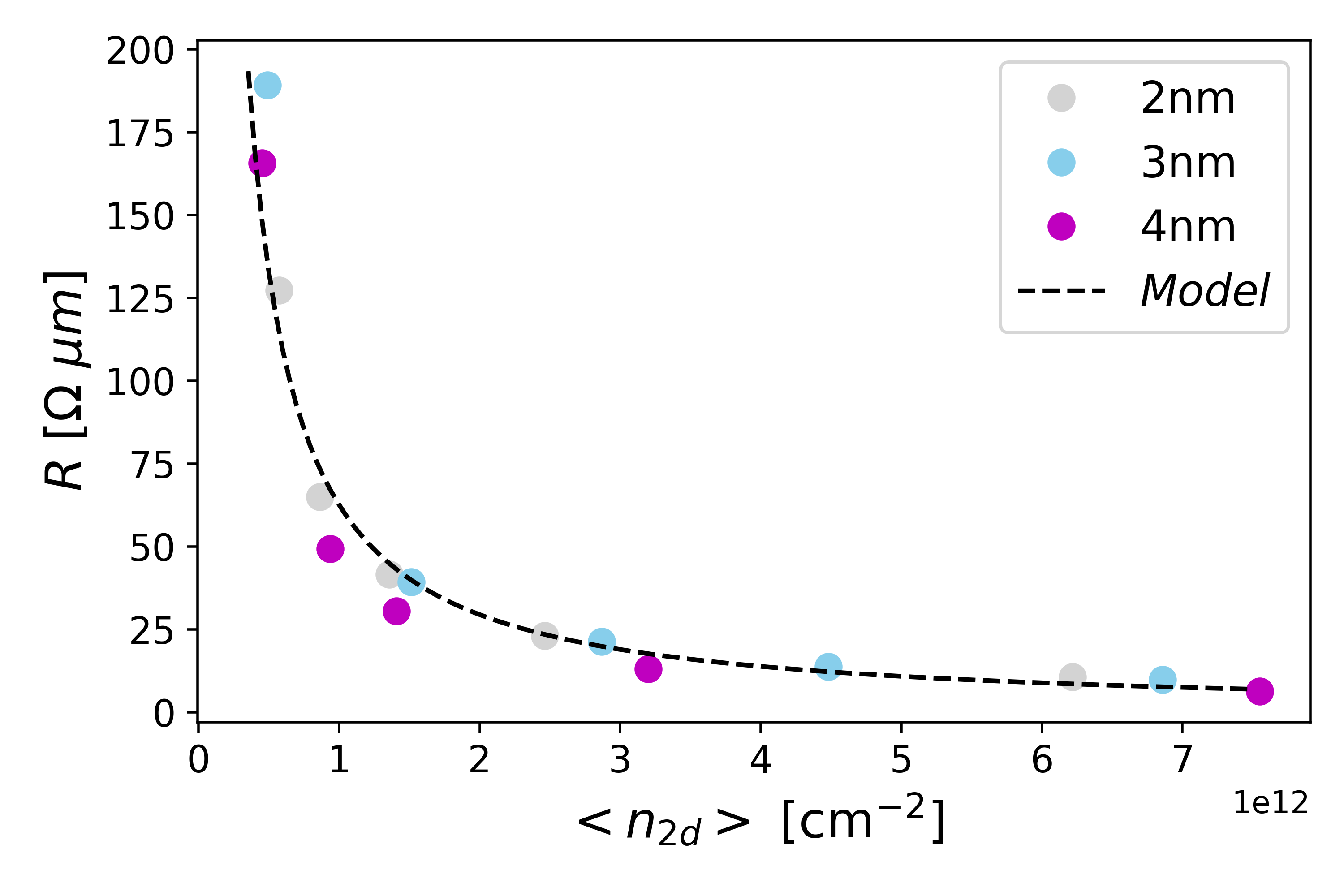}
    \end{center}
    \caption{Electrical resistance of the SiNW model in Fig.~\ref{fig:Device} as a function of the average carrier density for various nanowire diameters.}
    \label{fig:Resistance}
\end{figure}

As an application of the proposed method, we investigate the electrical resistance $R$ between SiNWs and metal contacts as a function of the carrier density $\langle n_{2d}\rangle$ for various diameters. We consider three SiNWs models with $2, 3$ and $4$ nm diameters. Fig.~\ref{fig:Resistance} shows that $R$ i) exhibits a clear $\langle n_{2d} \rangle ^{-\alpha}$ dependence (with $\alpha=1.068$) and ii) it is weakly affected by the diameter. This confirms that $R$ is dominated by the quasi-Fermi level drop near the contact regions~\cite{rideau2014experimental}. 

\section{CONCLUSIONS} 
We developed an algorithm to efficiently compute the self-energy matrices of large quasi-1D channels without sacrificing accuracy. By partitioning the contacts into PLs, we are able to significantly reduce the simulation time and to evaluate the electrical resistance of SiNWs with metallic contacts. This algorithm is key to simulate more realistic nanostructures with bulk-like contacts.

\newpage

%\section*{References}
%[1] L. Bourdet et al., \textit{J. Appl. Phys.} 119, 084503 (2016). 

%[2] M. L. Sancho et al., \textit{J. Phys. F} 14, 1205 (1984). 

%[3] K.S. Thygesen et al., \textit{Phys. Review B} 73, 035309 (2006). 

%[4] S. E. Mohney et al., \textit{Solid-State Electron.} 49, 227-232 (2005). 

%[5] J.Enkovaara et al., \textit{J. Phys.} 22, 253202 (2010). 

%[6] A. R. Dmitry et al., \textit{Theory of Quantum Transport at Nanoscale: An Introduction}. 

%[7] D. Rideau et al., \textit{SISPAD}, 101-104 (2014). 

%%%%%%%%%%%%%%%%%%%%%%%%%%%%%%%%%%%%%%%%%%%%%%%%%%%%%%%%%%%%%%%%%%%%%%%%%%%%%%%%
\newpage

\clearpage

\bibliographystyle{ieeetr}
\bibliography{bibliography}
%\begin{spacing}{.5}
%\bibliography{bibliography}
%\end{spacing}

%\section{Acknowledgements}

%\clearpage

\end{document}